# High-Temperature Superconductivity of the Fe-Se-H compound

S.I. Bondarenko[1], A.A. Prokhorov[2], N.N. Galtsov[1],
V.P. Timofeev[1], V.P. Koverya[1], I.S. Bondarenko[1], A.V. Krevsun[1]

[1] B. Verkin Institute for Low Temperature Physics and Engineering of the National Academy of Sciences of Ukraine, Kharkiv 61103, Ukraine
[2] Institute of Physics of the Czech Academy of Sciences, Prague, Czech Republic

**Abstract.**
Using an EPR spectrometer, the dependences of microwave power absorption on magnetic field up to 6000 Oe were measured at temperatures of 3.6–25 K and 295 K in a powdered compound Fe-Se-H, obtained by thermal diffusion of hydrogen into a FeSe single crystal with a critical temperature of 8 K. The shape of the dependences and the analysis of the measurement results confirm the superconductivity of the Fe-Se-H compound at normal pressure, both at temperatures of 3.6–25 K and at 295 K, i.e., at room temperature, equal to 22 $^0$C.

## 1. Introduction

Increasing the critical temperature ($T_c$) of superconductors is important scientific and applied problem. One way to address this is by synthesizing chemical elements with hydrogen [1, 2, 3, 4, 5]. In recent years, the highest critical temperatures ($T_c$=205-250 K) have been achieved in hydrides under high pressures of tens and hundreds of GPa [6, 7]. Moreover, hydrogen-doped superconductors obtained at normal atmospheric pressure have the highest $T_c$ of approximately 50 K [8, 9]. For widespread use of superconducting devices, they must operate at normal atmospheric pressure and have a critical temperature at room temperature, i.e., conventionally higher than 291 K (18 $^0$C). An alternative way to improve the properties of superconductors under normal external pressure may be to create internal, chemical pressure within them. During our studies of the effects of hydrogen, we discovered [10] that thermal diffusion of hydrogen into single crystals of the $FeTe_{0.65}Se_{0.35}$ compound leads to strong volumetric compression of its crystal lattice (up to 15%) and the generation of high internal pressure. As a result of thermal diffusion, the superconducting critical current density in this single crystal increased by a factor of 30, and $T_c$ increased by one degree [11]. Given this result, the aim of this study was to determine the effect of hydrogen thermal diffusion on FeSe, the base compound of this family of iron-based superconductors.

## 2. Experimental setup

Thermal diffusion of hydrogen into a Fe-Se single crystal with dimensions of approximately 2×2×0.4 mm$^3$ was performed in a steel chamber for 10 hours at a hydrogen temperature of 180 $^0$C and a hydrogen pressure of approximately 5 atm. After cooling and opening the chamber, it was discovered that the crystal had lost stability and had transformed into a powder of particles ranging in size about 200 microns. A possible cause of the crystal destruction is strong internal compression, similar to the compression of a $FeTe_{0.65}Se_{0.35}$ single crystal as a result of hydrogen thermal diffusion. The change in the FeSe crystal's properties as a result of hydrogen incorporation led to the formation of a new compound, Fe-Se-H. To determine the superconducting properties of the Fe-Se-H compound by a non-contact method, the powder was placed in the microwave resonator of an EPR spectrometer [12]. The spectrometer's output signal is the derivative with respect to the magnetic field (d$P$/d$H$) of the microwave power ($P$) absorbed by the powder. The signal was measured as a function of the external magnetic field ($H$) up to 6000 Oe and temperature ($T$) from 3.6 K to 295 K (22 $^0$C). The absorption of microwave power in the powder particles occurred in a resonant and non-resonant manner at a radiation frequency of 9.4 GHz, which is the frequency of the Bruker X-/Q-band E580 FT/CWELEXSYS spectrometer used in the studies. The superconducting properties of the powder were determined based on the analysis of non-resonant EPR signals at temperatures of 3.6 K – 25 K and at 295 K. The absorption of microwave power in type II superconductors in a mixed state is usually associated [13] with the flow of microwave current through those regions of it that are occupied by Abrikosov vortices. It is in the central normal parts of the vortices (cortices) [14] that microwave power dissipation occurs. The larger the sample area occupied by vortices, the greater the power absorbed. The vortex density in the sample increases as the constant magnetic field exceeds the first critical value. Due to changes in the vortex density and their interaction, the magnetic-field dependences of the non-resonant EPR signal from the sample are nonlinear. The existence of such nonlinear magnetic-field dependences indicates superconductivity of the sample and its belonging to the class of type II superconductors [15].

## 3. Experimental results and discussion

### 3.1 Magnetic-field dependences of EPR signals from a FeSe single crystal and a Fe-Se-H compound in a magnetic field of 0-6000 Oe

Figures 1 and 2 show, for comparison, the magnetic-field dependences of the signal (MFDS) of the spectrometer from the initial FeSe single crystal, previously obtained by us [15], and the MFDS from the Fe-Se-H compound.

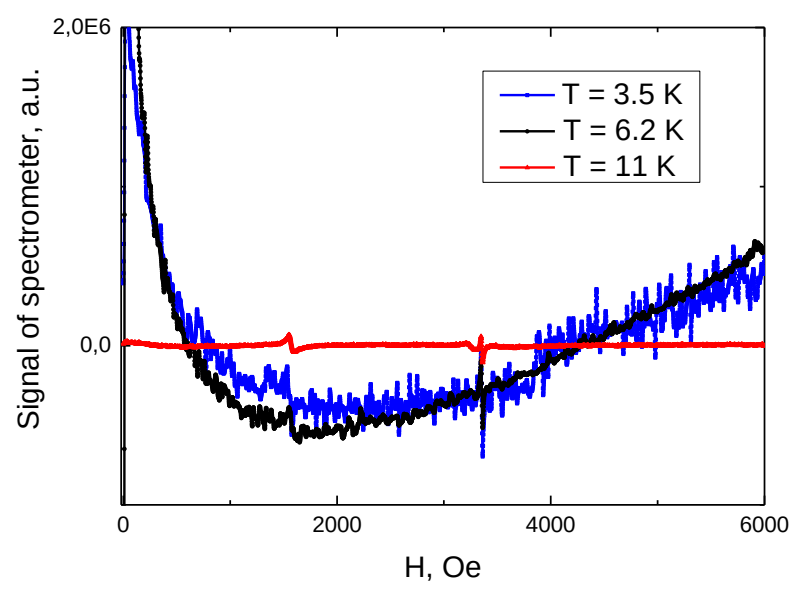

Fig. 1. MFDS from a single crystal FeSe at temperatures 3.5 K, 6.2 K, 11 K.

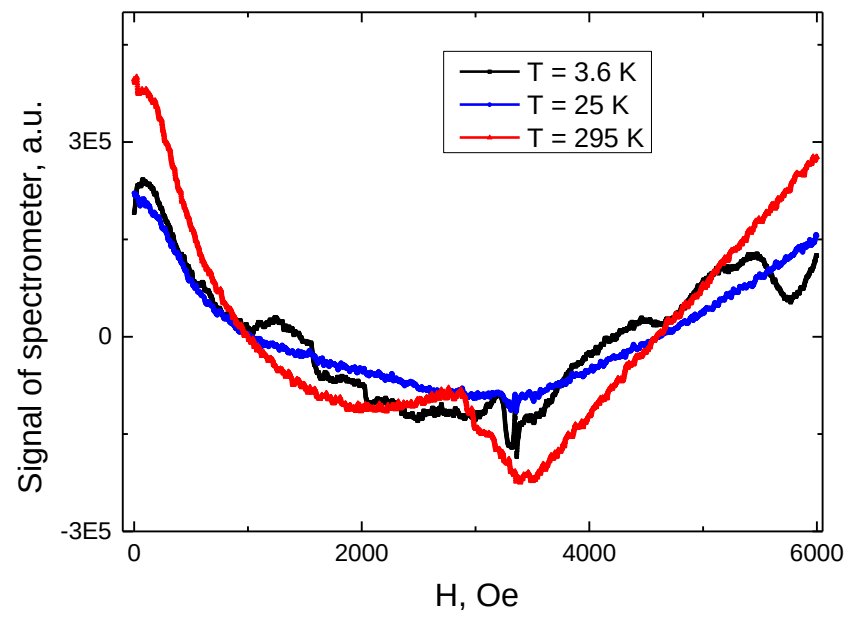

Fig. 2. MFDS from the compound Fe-Se-H at temperatures 3.6 K, 25 K, 295 K.

The original FeSe single crystal is a type-II superconductor. The typical shape of its U-shaped MFDS (Fig. 1) corresponds to its mixed superconducting state [15]. These dependences have minima in the region of a magnetic field of about 1300 Oe with two branches to the left and right of this minimum. Considering that the dependences are the first field derivatives (d*P*/d*H*) of the field dependences of microwave power absorption *P* (*H*), it can be assumed that in the region of the minimum there is an inflection of the *P*(*H*) dependences, where a change in the power absorption mechanisms occurs. It is likely that to the left of the inflection, the area of the superconducting region of the crystal without vortices exceeds the area occupied by vortices, and vice versa to the right. In the inflection region, the areas are equal. At temperatures above $T_c$=8 K (*T*=11 K), the crystal is in a normal state, and its EPR signal is independent of the magnetic field up to 6000 Oe. Furthermore, the dependence at *T*=11 K shows two resonant EPR signals: a broader one at 1400 Oe and a narrower one at 3400 Oe, corresponding to *g*-factor values of 4.2 and 2.0, respectively [15].

Figure 2 shows the nonlinear U-shaped EPR spectra of the powder from a magnetic field up to 6000 Oe at temperatures from 3.6 K to 25 K, as well as at a temperature of 295 K. These dependences are similar to the U-shaped dependences of the superconducting single crystal in Figure 1.

Figure 3a and b show the d*P*/d*H* (*H*) and *P*(*H*) dependences for the powder at *T*=295 K. The *P* (*H*) dependence was obtained by integrating the d*P*/d*H* (*H*) dependence. It can be seen that the d*P*/d*H* (*H*) dependence also has a minimum, but at a field of 3400 Oe, while the *P*(*H*) dependence has an inflection point at this field. From the *P* (*H*) dependence, it follows that microwave power absorption in the powder sample increases nonlinearly as the field increases from zero to 6000 Oe. This process in the powder is similar to the change in the EPR signal in a FeSe single crystal and can be explained by a similar change in vortex density.

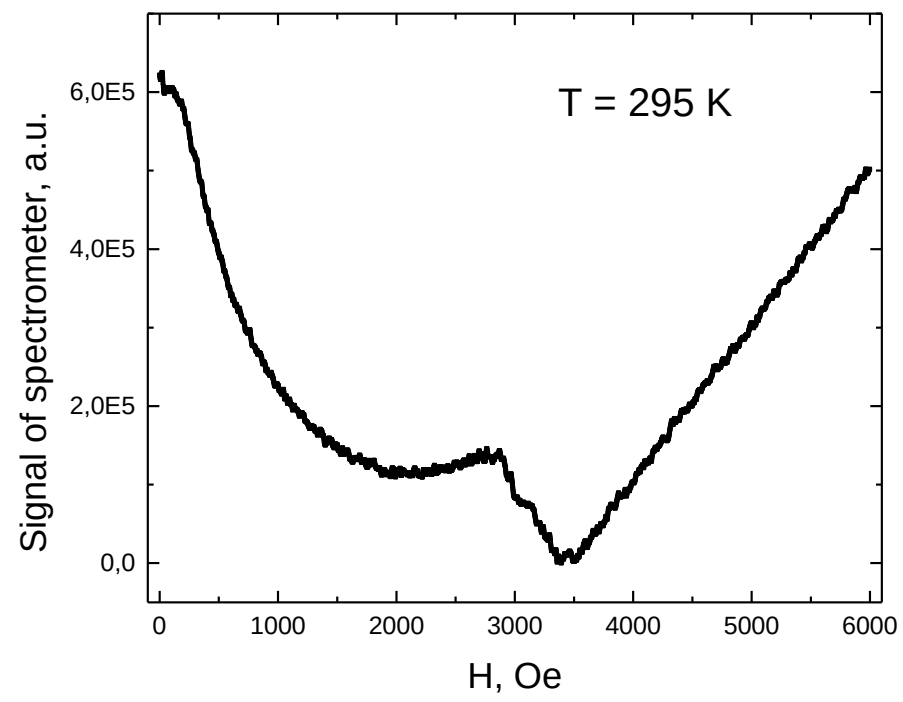

a)

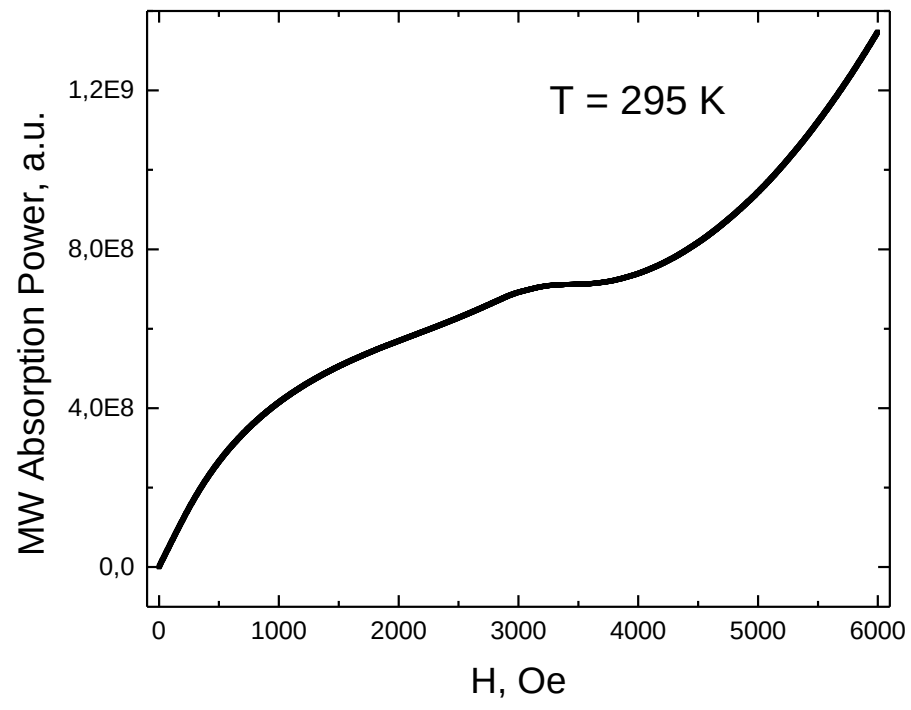

b)

Fig. 3. Dependences d*P*/d*H* (*H*) (a) and *P*(*H*) (b) for powder at *T*=295 K.

The presence of U-shaped MFDS powder, typical for a superconducting mixed state, is the basis for believing that the Fe-Se-H compound is also in a superconducting mixed state at temperatures of 3.6 K to 25 K and at 295 K. The preservation of the U-shaped shape of the MFDS powder field at a temperature of 295 K indicates that the $T_c$ of the powder is above 295 K.

At the same time, the powder's MFDS exhibits several features that distinguish it from the MFDS of a FeSe single crystal. These include: the powder's MFDS minimum is at 3400 Oe, there is no resonance at 1300 Oe, and the resonance at 3400 Oe at $T$ = 295 K is greatly broadened. These features indicate that the powder's electronic state differs from that of a single crystal.

## 3.2 Magnetic field dependences of signals (MFDS) of the Fe-Se-H compound in a weak magnetic field

MFDS in a weak magnetic field provide a clearer understanding of the initial sections of the dependences in Figures 1 and 2. Figures 4 and 5 show the MFDS of the powder and single crystal in the field range from zero to 30 Oe at different temperatures.

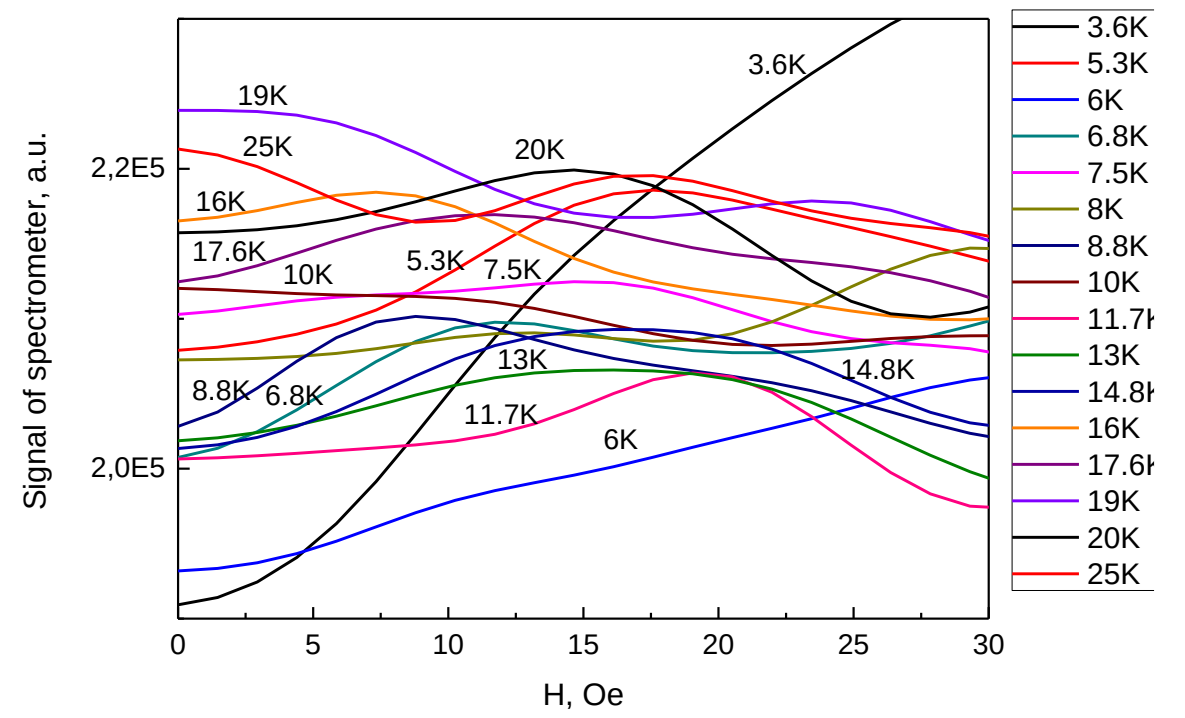

Fig. 4. MFDS of the Fe-Se-H compound at zero and weak magnetic field up to 30 Oe at $T$=3.6-25 K.

Fig. 5. MFDS of a FeSe single crystal at $T$ = 3.5-7.8 K [15].

The features of the crystal and powder MFDS in Figs. 1 and 2 are poorly distinguishable in a weak magnetic field near zero. To determine the shape of these features, MFDS measurements were conducted in a weak magnetic field up to 30 Oe. Figure 4 shows the nonlinear MFDS of the Fe-Se-H compound at temperatures of 3.6-25 K, and Figure 5 shows the nonlinear MFDS of an FeSe single crystal in the same weak magnetic field range at temperatures from 3.5 K to 8 K. Figure 5 shows that the dependences have both positive and negative values. In Fig. 4, all values are positive. As the crystal temperature approaches the critical value (8 K), the MFDS amplitude of the superconducting crystal decreases to zero, and the dependences degenerate into a straight line in Figure 5, while the signal for the Fe-Se-H compound remains significant and nonlinear even at 25 K. Thus, the existence of magnetic-field dependences of the non-resonant EPR signal from the studied samples is evidence of their superconductivity. Besides, properties of the MFDS powder indicate, firstly, that there is a non-superconducting oxide on the surface of the powder particles and, secondly, that the powder is superconductor at temperatures three times higher than the crystal's $T_c$. The latter was one of the motivations for taking EPR measurements of the powder not only at temperatures up to 25 K but also at 295 K.

### 3.3 Quasi-periodic EPR signal dynamics of the Fe-Se-H compound at *T* = 295 K

Figure 2 shows the EPR signal dynamics of Fe-Se-H powder over a wide range of magnetic fields, from zero to 6000 Oe, at temperatures of 3.6 K to 25 K and at 295 K. Important details of these dynamics were revealed by recording the EPR signal dynamics at several points in the curves in Figure 2 with higher field resolution. This increased resolution was achieved by narrowing the field range (to 30 and 120 Oe) in which the typical quasi-periodic change in the EPR signal occurred against the background of higher field values up to 6000 Oe. Some of them are shown in Figs. 6-9. These figures demonstrate that the EPR signal is quasi-periodic and has a characteristic period close to 20 Oe.

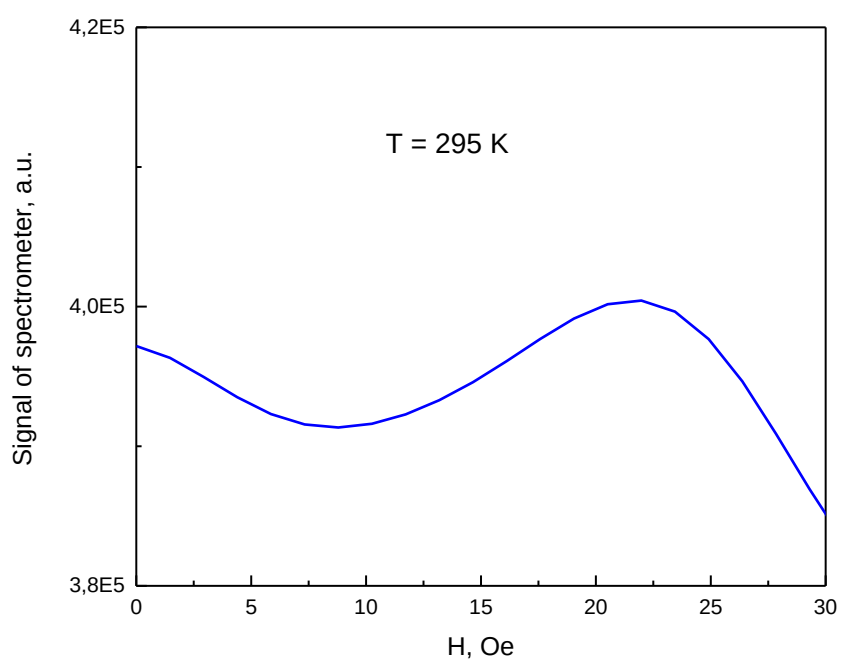

Fig. 6. MFDS powder in a field of 0-30 Oe.

Fig. 7. MFDS powder in a field of 0-120 Oe.

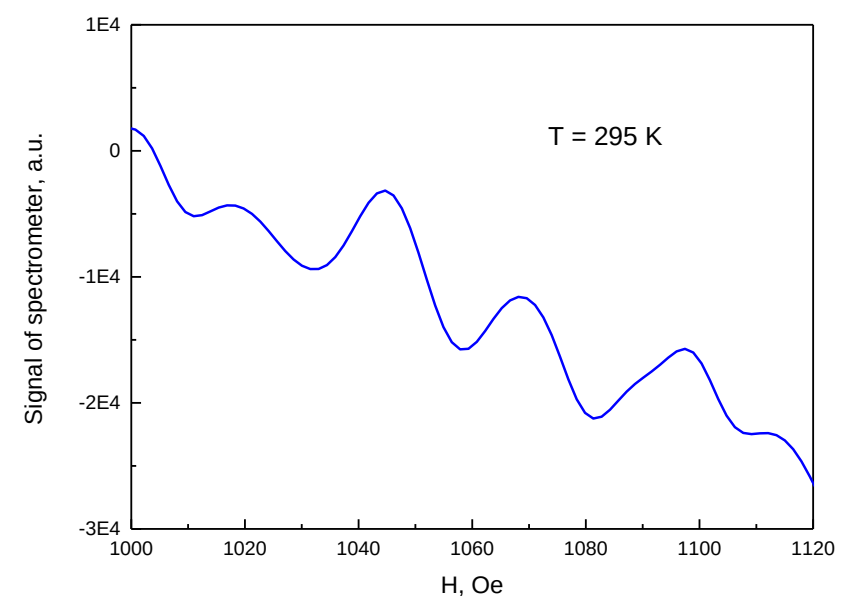

Fig. 8. MFDS powder in a field of 1000-1120 Oe.

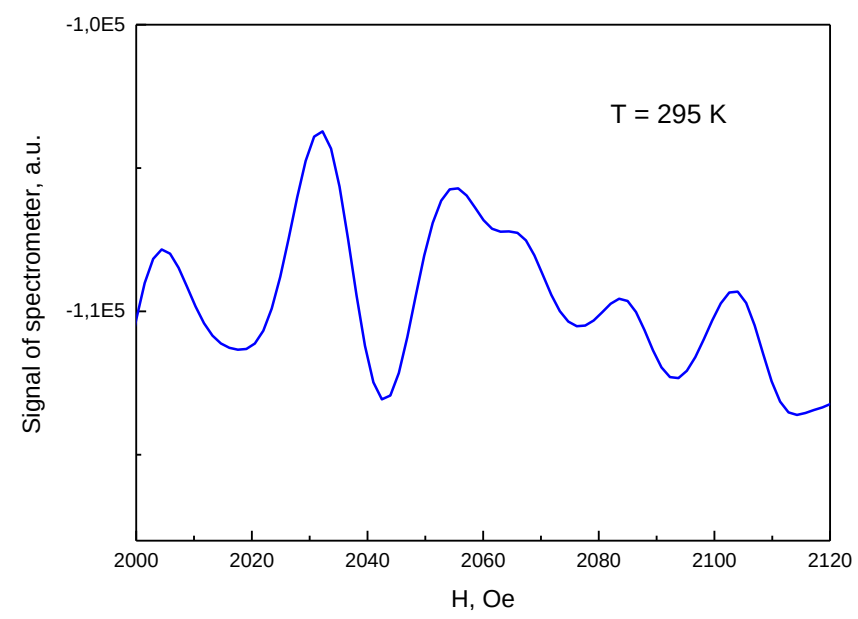

Fig. 9. MFDS powder in a field of 2000-2120 Oe.

It should be noted that high-resolution EPR signals for a FeSe crystal in a mixed state are also characterized by quasi-periodic EPR signals with an average period value close to 20 Oe. As an example, two dependences are shown at *T*=7 K (Figs. 10, 11).

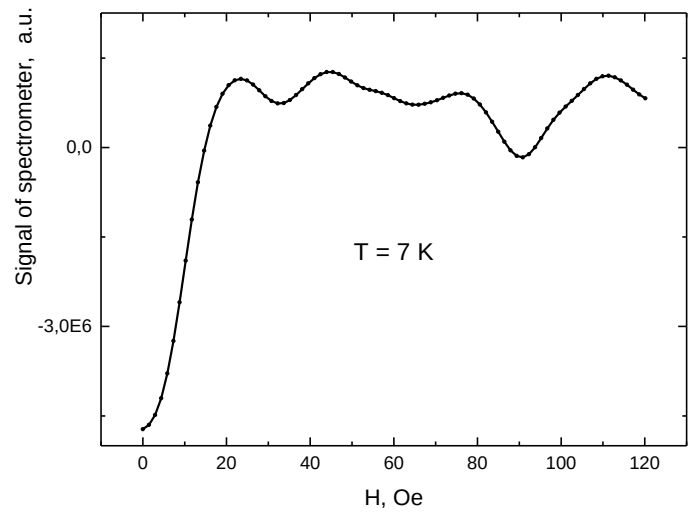

Fig. 10. MFDS of a FeSe crystal in a field of 0-120 Oe.

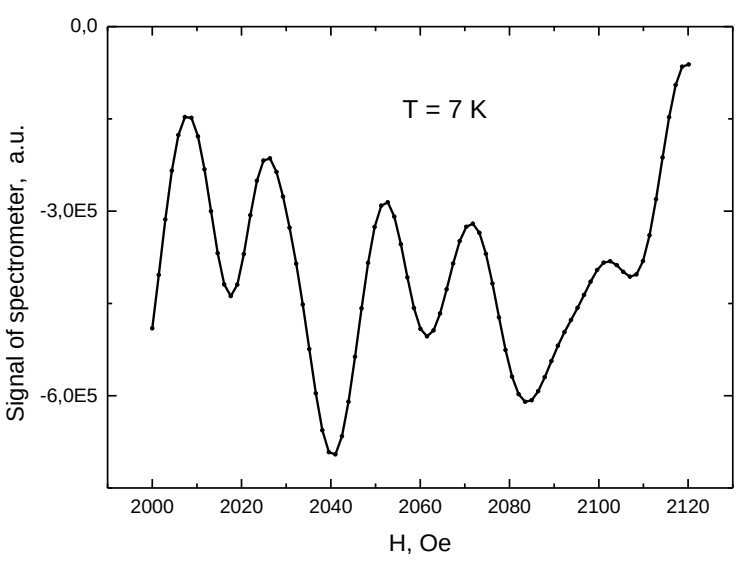

Fig. 11. MFDS of a FeSe crystal in a field of 2000-2120 Oe

Thus, quasi-periodic variations in the EPR signal of the Fe-Se-H compound with a period of about 20 Oe exist both at weak fields of about 30 Oe and at strong fields of up to 6000 Oe, which can be caused by the presence of Abrikosov vortices, both in the superconducting single crystal at temperatures below $T_c$ = 8 K and in the Fe-Se-H compound at $T$ = 295 K. The quasi-periodic variations in the signal when increasing the field can be explained by the presence of discrete absorption of microwave power in the Abrikosov vortices located in the particles of the compound. Some scatter in the amplitude and period of the EPR signals at different values of the external field $H$ can be explained by the simultaneous entry of a bundle of vortices into the particles, a difference in the orientation of some particles from the direction of the external measuring magnetic field, and a change in the interaction between the vortices as the external field increases.

Each Abrikosov current vortex generates a magnetic flux equal to the flux quantum $\Phi_0$. Given the similarity between the experimentally observed values of the period $\delta H$ of the EPR signal change for a single crystal at $T$=7 K ($T/T_c$=0.87) and a compound at 295 K, its value can be calculated from the relation:

$$\Phi_0 = \mu_0\, \delta H\, S_0\,, \qquad (1)$$

where $\mu_0$ and $S_0$ are the magnetic susceptibility of a vacuum ($4\pi 10^{-7}$ H/m) and the area of the current vortex generating the magnetic flux quantum $\Phi_0$. If we take into account that the diameter ($d$) of the vortex is close to the London penetration depth of the magnetic field $\lambda(T)$ [14] into the FeSe single crystal, then its area $S_0$ is equal to $\pi\, d^2/4 \approx \pi\, [\lambda(T)]^2/4$. Thus,

$$\Phi_0 \approx \mu_0\, \delta H\, \pi\, [\lambda(T)]^2/4. \qquad (2)$$

It is known [16] that for a single FeSe crystal $\lambda(0) = 0.9\times10^{-6}$ m, and $\lambda(7\text{ K}) \approx 10^{-6}$ m. After substituting $\lambda^2\,(7\text{ K}) = 10^{-12}$ m$^2$ and $\Phi_0 = 2\times10^{-15}$ Wb into (2), we obtain:

$$\delta H \approx \Phi_0\, /\mu_0\, \pi\, [\lambda(T)]^2/4 \approx 2\times10^3 \text{ A/m} = 22.5 \text{ Oe}. \qquad (3)$$

Thus, the calculated period is close to the measured value (20 Oe), and the quasi-periodic changes in the EPR signal with increasing external field shown in Figs. 6-11 correspond to the number of vortices $N$ that entered the powder particles with an increase in the external magnetic field by 20 $N$ Oe. Consequently, such changes in the EPR signal from the Fe-Se-H compound are evidence that it belongs to the type-II superconductors, and EPR spectrometry allows visualization of Abrikosov vortices in the superconductor. This is the second argument in favor of the existence of type-II superconductivity in the Fe-Se-H compound at $T$ = 295 K (22 $^0$C).

As can be seen from (2), the period of the signal change depends on the penetration depth of the magnetic field into the superconductor. The equality of the periods $\Delta H$ for the FeSe single crystal near $T_c$ ($T$=7 K) and the Fe-Se-H compound at $T$=295 K implies equality of penetration depths at equal relative temperatures ($T/T_c$). Since $T/T_c$ at 7 K is 0.87 for the single crystal, then the critical temperature of the Fe-Se-H compound at the same relative temperature should be 340 K (67$^0$ C). From the BCS relationship $\Delta\,(0)/kT_c = 1.76$ for the energy gap $\Delta(0)$ and the critical temperature $T_c$, it follows that for the Fe-Se-H compound $\Delta\,(0) = 8.6\times 10^{-21}$ J = 54 meV.

**Conclusions**

1. Superconductivity of an iron-based compound was obtained for the first time at room temperature ($22^0$ C). The existence of high-temperature superconductivity in the Fe-Se-H compound is substantiated by:
a) the existence of similar U-shaped magnetic-field dependences of the EPR signal of the Fe-Se-H compound at temperatures of 3.6–25 K and 295 K (22 $^0$C) and U-shaped magnetic-field dependences of the EPR signal of the superconducting FeSe single crystal at temperatures of 3.5–8 K, characteristic of the mixed superconducting state,
b) the possible EPR visualization of Abrikosov quantum vortices in Fe-Se-H compound particles at $T$ = 295 K.
2. It is shown that thermal diffusion of hydrogen into a FeSe single crystal enables superconductivity to be achieved in the Fe-Se-H compound at normal external pressure and a temperature 37 times higher than the critical temperature of the initial FeSe single crystal (8 K). The probable cause of high-temperature superconductivity in the Fe-Se-H compound is significant internal compression of the crystal lattice during the creation of chemical bonds between hydrogen and iron and selenium atoms, with the simultaneous emergence of a new electronic state.
3. The calculated critical temperature of the Fe-Se-H superconductor is 340 K (67 $^0$C), which corresponds to an energy gap at zero temperature of 52 meV.
4. Non-resonant EPR spectroscopy enables the visualization of quantum Abrikosov vortices in magnetic fields ranging from the first critical value to 6000 Oe at temperature to 295 K. This property expands the possibilities for studying the dynamics of Abrikosov vortices, for the development of new superconductors, and expands the application of superconductivity by eliminating the need for cooling superconducting devices.

The authors express their gratitude to D. A. Chareev for preparing single crystals FeSe, M. I. Kobets, S. N. Shevchenko for useful discussions, and A. V. Dolbin, A. L. Solovjov for assistance in organizing the experiments.

This work was financially supported by the of leading Program of the National Academy of Sciences of Ukraine “Fundamental research on the most important problems of natural sciences” (section "Quantum nano-sized superconducting systems: theory, experiment, practical implementation"). State registration number of the work is 0122U001503. V.P.K. was supported by grant of the IEEE Magnetic Society (project No. 9918) and National Research Foundation of Ukraine (Grant No. 2025.07/0044).